\documentclass{article}
\usepackage{spconf,amsmath,graphicx}
\usepackage{booktabs}
\usepackage{afterpage}
\usepackage{hyperref}

\title{TIA: A Teaching Intonation Assessment Dataset in Real Teaching Situations}
\name{Shuhua Liu, Chunyu Zhang, Binshuai Li, Niantong Qin, Huanting Cheng, Huayu Zhang\sthanks{Corresponding author: Huayu Zhang, zhanghy680@nenu.edu.cn}}
  \address{School of Information Science and Technology, Northeast Normal University, China}
\begin{document}
\topmargin=0mm
%
\maketitle
\begin{abstract}
Intonation is one of the important factors affecting the teaching language arts, so it is an urgent problem to be addressed by evaluating the teachers’ intonation through artificial intelligence technology. However, the lack of an intonation assessment dataset has hindered the development of the field. To this end, this paper constructs a Teaching Intonation Assessment (TIA) dataset for the first time in real teaching situations. This dataset covers 9 disciplines, 396 teachers, total of 11,444 utterance samples with a length of 15 seconds. In order to test the validity of the dataset, this paper proposes a teaching intonation assessment model (TIAM) based on low-level and deep-level features of speech. The experimental results show that TIAM based on the dataset constructed in this paper is basically consistent with the results of manual evaluation, and the results are better than the baseline models, which proves the effectiveness of the evaluation model.
\end{abstract}
\begin{keywords}
Teaching Intonation Assessment Dataset, Teaching Intonation Assessment Model, Wav2vec2.0, Bi-LSTM, Attention Mechanism
\end{keywords}
\afterpage{%
  \begin{figure*}[t!]
    \centering
    \includegraphics[height=0.20\textheight]{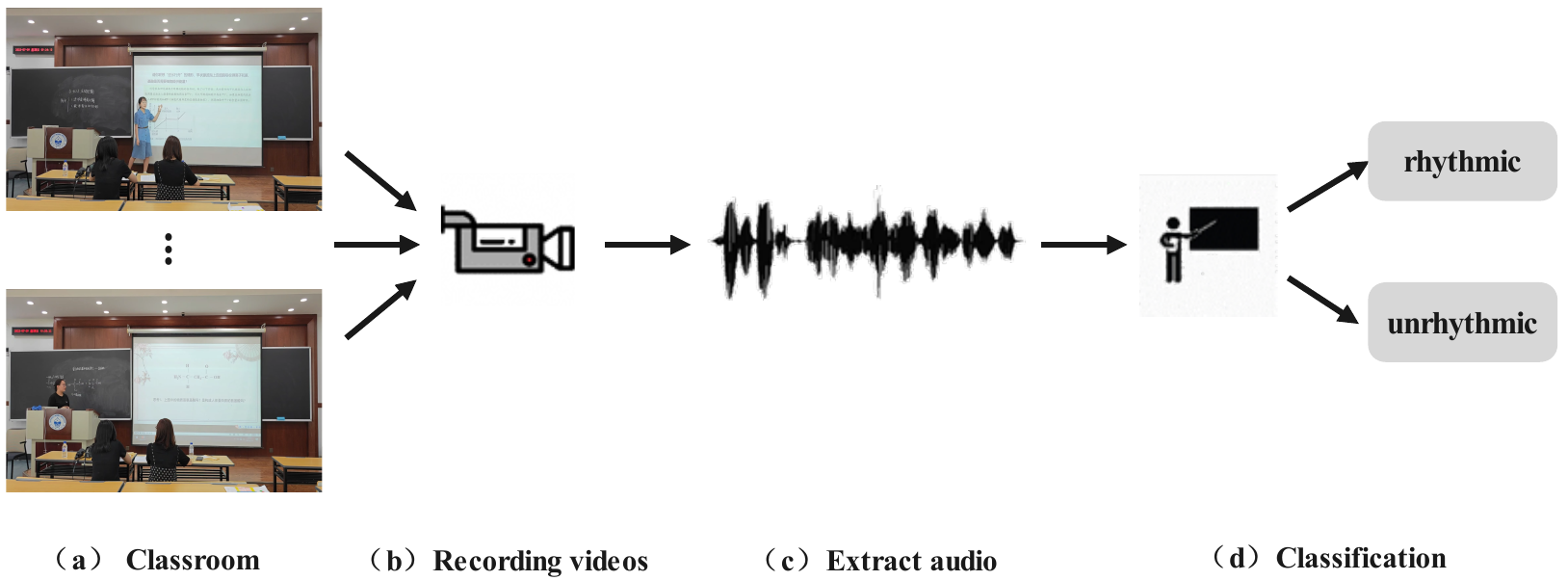}
    \caption{Overview of the acquisition process and content of the TIA dataset. Each recorded teacher speech contains audio and intonation labels provided by educational experts.}
    \label{fig:figure1}
  \end{figure*}
}

\section{Introduction}
\label{sec:intro}

Voice is a basic tool for human communication. However, the teaching language in the classroom has strict requirements for pitch, melody, and loudness, etc. The appropriate intonation can greatly improve the teaching effect. Teaching is a bilateral activity consisting of the teaching of teachers and the learning of students. Therefore, the rhythm of the teaching language should be adapted to various factors such as students, teaching content, teaching environment, and teaching requirements. The tone of the voice, the change of high and low, fast and slow, are the language skills commonly used by teachers in classroom teaching. Reasonable mastery of tone can effectively mobilize students' interest in learning, focus students' attention, and improve teaching effects. Therefore, this study will explore the assessment standard of the teaching intonation based on artificial intelligent technology. To end this, we firstly construct a teaching intonation assessment dataset (TIA). As far as we know, TIA is the first teaching intonation assessment dataset. It covers 9 disciplines, with rich disciplinary characteristics. In order to evaluate the TIA dataset, this study proposes a teaching intonation assessment model (TIAM). This model extracts the low-level features and deep-level feature of audio signals, and then fuses these features by attention mechanism. Finally, the assessment result ``rhythmic” or ``unrhythmic” is obtained. The experimental results show that the effectiveness of the proposed TIAM.

The main contributions of this paper are listed as follows:

(1) This paper constructs a teaching intonation assessment (TIA) dataset. It provides a new benchmark dataset for teaching intonation assessment. This dataset is unique in its characteristics of discipline richness, data diversity, and real-world situations. The construction of this dataset can promote further research and development of teaching intonation assessment algorithms.

(2) A teaching intonation assessment model (TIAM) is proposed. TIAM consists of two branches, one is utterance-level branch which extracts features from 15 seconds audio segment, and the other is low-level branch which extracts low-level features from 25 ms audio segment. Finally, TIAM fuses these features by the attention mechanism for classification.

(3) A large number of experiments are carried out on TIAM datasets. The experiments show that the results of teaching intonation assessment based on TIA dataset are basically consistent with those of manual assessment, and the proposed teaching intonation assessment model (TIAM) is superior to the baseline models.

\afterpage{%
  \begin{figure*}[t!]
    \centering
    \includegraphics[height=0.31\textheight]{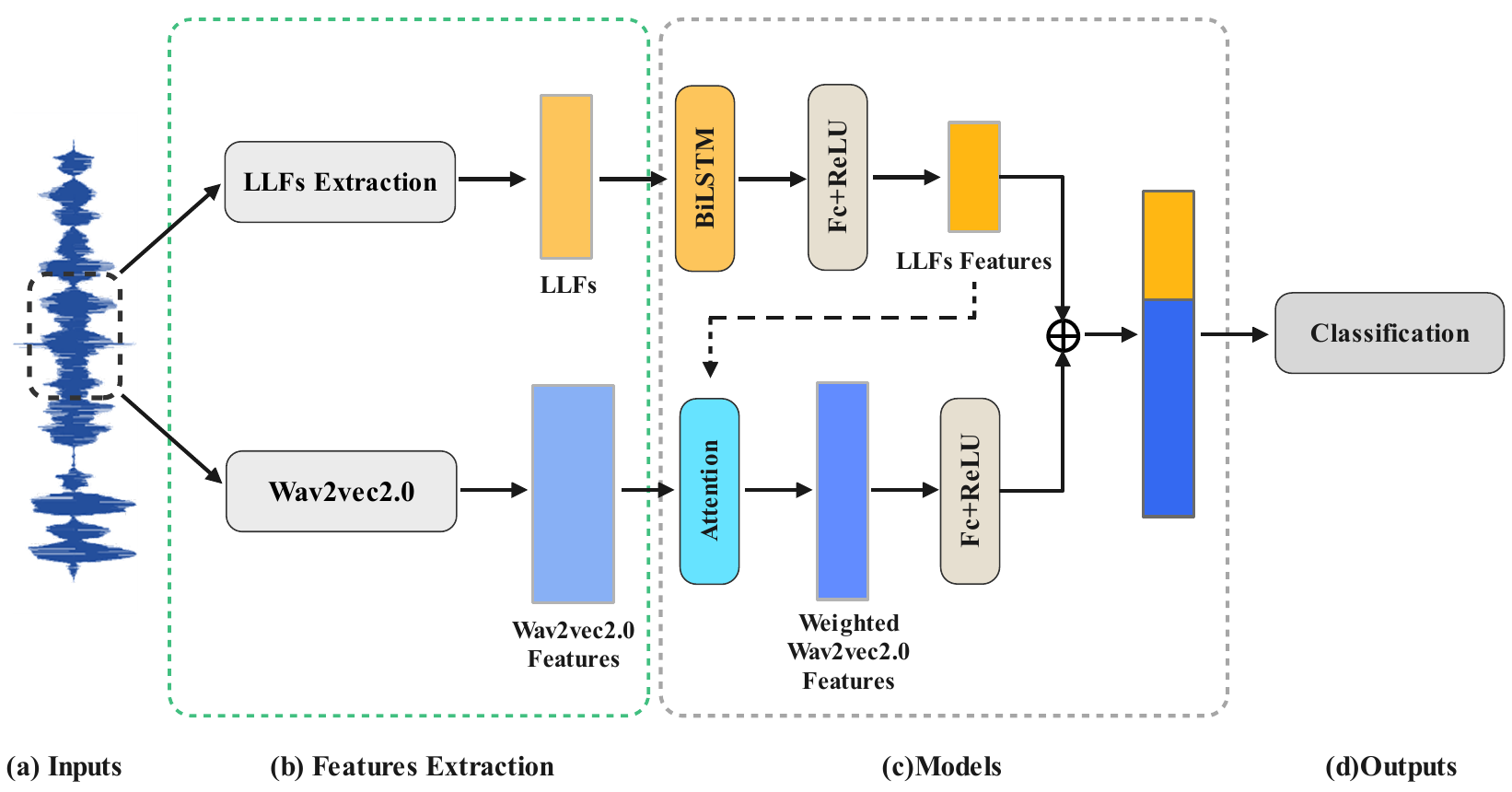}
    \caption{The overall architecture of our proposed method TIAM.}
    \label{fig:figure2}
  \end{figure*}
	\begin{table*}[ht]
	\centering
	\caption{Distribution of Classes by Discipline in the TIA Dataset}
	\label{tabelsubject}
	\begin{tabular}{c|cccccccccc}
	\hline\hline
	\textbf{Discipline} & Chinese & Maths & English & Physics & Chemistry & Biology & Politics & History & Geography & \textbf{Total}\\
	\textbf{Number} & 111 & 49 & 58 & 34 & 24 & 29 & 16 & 42 & 33 & 396\\
	\hline\hline
	\end{tabular}
	\end{table*}
}

\section{Related work}
\label{sec:related}

\begin{figure}[t!]
\centering
\includegraphics[height=0.22\textheight]{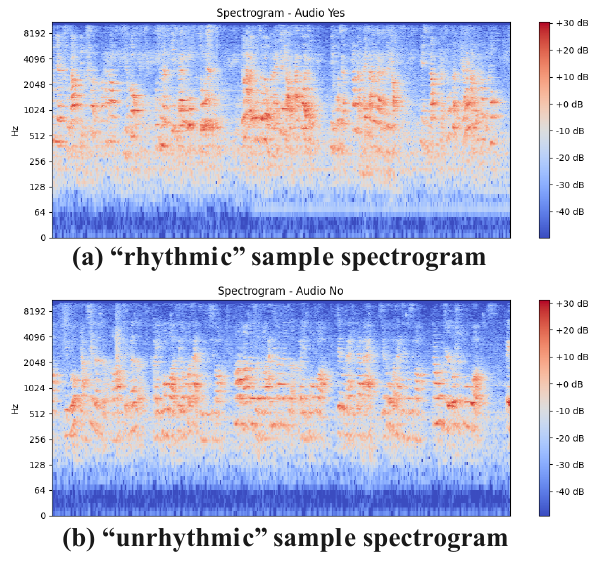}
\caption{Spectrogram samples in the TIA dataset}
\label{fig:figure3}
\end{figure}

Soviet educator Makarenko once said: ``The language of instruction is the most important means of teaching." Teachers pay attention to the teaching language art in classroom teaching, which will inevitably increase the attractiveness of teaching. Among them, tone and intonation are important parts of the teaching language. Tone is the changing trend of pitch with time. In Chinese, tone plays an essential role for distinguishing meaning. Therefore, some research has carried out base on machine learning for Mandarin tone recognition. In early research, traditional machine learning approaches are adopted, such as distributed HMM\cite{1} , SVM\cite{2} , and BP\cite{3} network to recognize tone. Yan et al.\cite{4} proposed a Mandarin machine learning method based on random forest and feature fusion, where corresponding tone classifiers are modeled and optimized. With the emergence of deep learning, some tone recognition approaches have been proposed. Ryant et al.\cite{5}  used a deep neural network (DNN)-based classifier to identify five tone categories in Mandarin broadcast news based on 40 Mel frequency cepstral coefficients (MFCC). Tan et al.\cite{6} proposed DNNHMM for Mandarin speech recognition and Lin et al.\cite{7} proposed an improving method for mandarin tone recognition based on DNN. Compared with mandarin, Chinese dialect’s tone is more complex, so Zhang et al.\cite{8} used gated spiking neural P systems for Chinese dialect tone’s recognition. Loren Lugosch et al.\cite{9} propose a method for continuous speech tone recognition in tonal languages using Convolutional Neural Networks (CNN) and Connectionist Temporal Classification (CTC). Yang et al.\cite{10} explored the use of bidirectional long-short term memory (BiLSTM) with an attention mechanism for Mandarin tone recognition, aiming to process tone changes in continuous speech.

Summarily, above research focuses on tone recognition, that is, if the pronunciation is accurate. However, there is little research on intonation. S. Wager et al.\cite{11} proposed an ``Intonation" dataset of amateur vocal performances with a tendency for good intonation, collected from Smule, Inc. The dataset can be used for music information retrieval tasks. Regarding the recognition of teachers’ intonation in class, with our knowledge, there are no reports of relevant studies and datasets. Therefore, this study will firstly construct a dataset for teaching intonation assessment and then evaluate the dataset on a proposed model.

\section{The construction of TIA Dataset}
\label{sec:TIA construction}
In order to fill the gap in the teaching intonation assessment dataset and to facilitate a more accurate identification of teacher intonation on the classroom, this paper proposes the teaching intonation assessment dataset TIA. Figure \ref{fig:figure1} shows an overview of the acquisition process and content of the TIA dataset.

TIA consists of 396 real classroom lectures by 396 teachers. As shown in Table \ref{tabelsubject}, it covers 9 disciplines such as Chinese, Maths, English, Physics, Chemistry, Biology, Politics, History, and Geography. We intercept 48 hours of teachers’ lecture audios from over 100 hours of real classroom recordings, and then classify these audios into two categories based on whether the teacher’s tone is rhythmic or not by expert evaluation manually.

Table \ref{tabelyesorno}  shows the distribution of class labels for the 11,444 15-second, 16 kHz audio samples in the TIA dataset , with 8,507 ``rhythmic" labels and 2,937 ``unrhythmic" labels. We randomly selected a ``rhythmic" sample and a ``unrhythmic" sample from the dataset and generated spectrograms for them. The spectrograms of ``rhythmic" sample and ``unrhythmic" sample are shown in Fig. \ref{fig:figure3} (a) and (b).

The spectrogram of ``rhythmic" sample exhibits obvious frequency and intensity dynamics, with up and down fluctuations in multiple frequency bands, which reflects the obvious intonation changes in speech.

In contrast, the spectrograms of the ``unrhythmic" samples showed relative stability in frequency and intensity, with frequency distributions concentrated in the same region and with high persistence, which demonstrated no significant intonation changes in speech. Therefore, it is practical to use the TIA dataset for the speech intonation assessment task.

\begin{table}[ht!]
\centering
\caption{Distribution of Class Labels}
\label{tabelyesorno}
\begin{tabular}{c|c}
\hline\hline
\textbf{Class} & \textbf{Number of Labels} \\
\hline
rhythmic & 8,507 \\
unrhythmic & 2,937 \\
\hline
\textbf{Total} & 11,444\\
\hline\hline
\end{tabular}
\end{table}

\section{Methodology}
\label{sec:Methodology}
In this section, we describe the teaching intonation assessment model by combination of low-level and deep-level features of speech. Figure \ref{fig:figure2} shows the overall structure of our proposed method. As illustrated, after splitting the original audio utterances into several segments, the low-level features of the segments and the deep-level features extracted by wav2vec2.0\cite{wav2vec2} are introduced into their respective feature encoder networks and fused with the proposed attention mechanism\cite{att} for the final intonation assessment.

\subsection{Inputs and Features Extraction}
\label{ssec:Inputs and Features Extraction}
We denote the Speech low-level features (LLFs) and wav2vec2.0 features, which are obtained from the same audio segment $x$, as $x_l$ and $x_w$.
\subsection{Intonation Assessment Model}

\label{ssec:Intonation Assessment Model}

The original audio clip is processed by the wav2vec2.0 processor to obtain the target wav2vec2.0 output as
\begin{equation}
\label{eqn-2} 
  x_w=Wav2vec2.0(x)
\end{equation}

The LLFs are processed by a Bi-LSTM\cite{longstm} with tanh as the activation function, with a dropout of 0.1, and fed into a linear layer with ReLU as the activation function, to obtain
\begin{equation}
\label{eqn-1} 
  x_l'=f_l(BiLSTM(x_l))
\end{equation}

The attention weights are obtained from $x_l'$ after a linear layer with Softmax as the activation function as
\begin{equation}
\label{eqn-3} 
  x_{att}=f_{att}(x_l')
\end{equation}

The wav2vec2.0 output $x_w'$ is multiplied by the attention weight $x_{att}$ to obtain the weighted wav2vec2.0 embedding (W2E) vector $x_w''$. The weighted W2E vector $x_w''$ is input to the linear layer with ReLU as the activation function with a dropout of 0.5, obtaining the final W2E vector as
\begin{equation}
\begin{aligned}
\label{eqn-4} 
  x_w'=(x_{att}*x_w)\\
  x_w''=f_w(x_w')
\end{aligned}
\end{equation}

The final LLFs and weighted W2E are concatenated together and the predictions of the intonation assessment model are as follows
\begin{equation}
\label{eqn-5} 
  \hat{y}=f(x_l'\oplus x_w'')
\end{equation}

\subsection{Outputs}
\label{ssec:Outputs}
We use the common cross-entropy loss for intonation classification as
\begin{equation}
\label{eqn-6} 
  \mathcal L=\mathcal L_{ce}(y-\hat{y})
\end{equation}

\section{Experiment}
\label{sec:expreiment}

\subsection{Experimental Setup}
\label{ssec:Experimentalsetup}
We used TIAM as a model for teaching intonation assessment, which uses low-level and deep-level features of speech in order to consider different levels of speech information. Low-level features such as MFCC, over-zero rate, pitch, and spectral centroid are extracted using the Librosa\cite{librosa} library and serve as key elements for assessing voice quality and intonation. Specifically, MFCC is robust against noise and captures voice timbre; over-zero rate monitors quick changes in speech; pitch gauges sound level; and spectral centroid is associated with loudness and timbre. Deep-level features are obtained from a pre-trained wav2vec2.0 transformer network, capturing time-domain characteristics of speech.

The hardware platform is NVIDIA GeForce RTX 4090 on Ubuntu 22.04.1. TIAM model uses the Pytorch framework. The optimizer for the model is AdamW with a learning rate of 5e-5. The training epochs is 300, batch size is 256. To avoid overfitting, the model use early-stopping strategy.

Our codes and TIA dataset will be available on Github\footnote{\href{https://github.com/zhangcy407/TIA}{https://github.com/zhangcy407/TIA}}.

\subsection{Results}
\label{ssec:results}
To the best of our knowledge, there are no deep learning models for intonation assessment datasets in recent years, so we chose to use a few SER models\cite{DST, joint, 9747095} as baselines. The experimental results in Table \ref{tabelpc} show that models with LLFs outperforms the assessment results without LLFs models, showing that LLFs of speech play an indispensable role in the intonation assessment task.
\begin{table}[ht!]
\centering
\caption{Performance comparison with SER model on TIA}
\label{tabelpc}
\begin{tabular}{c||ccc}
\hline\hline
\textbf{Model} & \textbf{Acc} & \textbf{F1} & \textbf{With LLFs}\\
\hline
DST\cite{DST} & 86.72 & 86.37 & N\\
Spectrum-Based\cite{joint} & 87.97 & 87.59 & N\\

Co-Attention\cite{9747095} & 88.27 & 87.98 & Y\\
Ours & \textbf{88.56} & \textbf{88.45} & Y\\
\hline\hline
\end{tabular}
\end{table}

\subsection{Ablation study}
\label{ssec:Ablation}

\begin{table}[ht!]
\centering
\caption{Ablation study}
\label{tabelas}
\begin{tabular}{c||cc}
\hline\hline
\textbf{Model} & \textbf{Acc} & \textbf{F1}\\
\hline
LLFs & 85.50 & 84.81 \\
W2E & 86.55 & 86.22 \\
LLFs+W2E & 87.07 & 86.90 \\
LLFs+W2E+Attention & \textbf{88.56} & \textbf{88.45} \\
\hline\hline
\end{tabular}
\end{table}
Our proposed method utilizes multiple levels of acoustic information in intonation assessment, which includes both time and frequency domain features. In order to study the effect of different combinations of acoustic information on the model performance, we conducted a series of ablation experiments and demonstrate them in Table \ref{tabelas}. The experimental results show that the performance of model  using W2E features are higher than using LLFs. This suggests that W2E features are more effective for expressing intonation information in intonation assessment tasks. In addition, we also try to use W2E and LLFs jointly and found that this combination can significantly improve the performance. This proves that using two features jointly can better capture the changing features of intonation and thus improve the assessment performance. In further experiments, we introduce the attention mechanism. The experimental results show that the recognition results of the model are further improved. It shows when fusing W2E features and LLFs, the attention mechanism can play an important role in helping the model to better focus on important information, thus further improving the recognition performance.

\section{Conclusion}
\label{sec:conclusion}
This study constructs a teaching intonation assessment dataset TIA based on some real classroom teaching recordings. To the best of our knowledge, this is the first teaching intonation assessment dataset. It covers 9 disciplines, 396 teachers, total of 11,444 utterance samples with a length of 15 seconds. This dataset has characteristics of discipline richness, data diversity, and real classroom situations, making it a high quality and challenging benchmark dataset. In order to test the validity of the dataset, this paper proposes a teaching intonation assessment model (TIAM) based on low-level and deep-level features of speech. The experimental results show the results are superior to the baseline models, which proves the effectiveness of the evaluation model. In future work, we will annotate the dataset with fine-grained labels, such as pitch, melody, loudness, and tone. For assessment results, we will set more grades instead of only two classes.

\section{Acknowledgement}
\label{sec:Acknowledgement}
This work is supported by the National Natural Science Foundation of China under the Grant 62277009, the project of Jilin Provincial Science and Technology Department under the Grant 20220201140GX.
\clearpage

\vfill\pagebreak

\bibliographystyle{IEEEbib}
\bibliography{bib}

\end{document}